\pdfoutput=1

\documentclass[11pt]{article}

\usepackage[final]{acl}

\usepackage{times}
\usepackage{latexsym}

\usepackage[T1]{fontenc}

\usepackage[utf8]{inputenc}

\usepackage{microtype}

\usepackage{inconsolata}

\usepackage{graphicx}

\title{ESC: Efficient Speech Coding with Cross-Scale Residual Vector Quantized Transformers}

\author{
 \textbf{Yuzhe Gu\textsuperscript{1,2}},
 \textbf{Enmao Diao\textsuperscript{2}}
\\
 \textsuperscript{1}University of Pennsylvania, Philadelphia, PA \\
 \textsuperscript{2}Duke University, Durham, NC
\\
 \texttt{tracygu@seas.upenn.edu} \quad
 \texttt{enmao.diao@duke.edu}
}

\begin{document}
\maketitle
\begin{abstract}
Neural speech codecs aim to compress input signals into minimal bits while maintaining content quality in a low-latency manner. However, existing neural codecs often trade model complexity for reconstruction performance. These codecs primarily use convolutional blocks for feature transformation, which are not inherently suited for capturing the local redundancies in speech signals. To compensate, they require either adversarial discriminators or a large number of model parameters to enhance audio quality. In response to these challenges, we introduce the \textbf{E}fficient \textbf{S}peech \textbf{C}odec (ESC)\footnote{Code and pretrained models available at \url{https://github.com/yzGuu830/efficient-speech-codec}}, a lightweight, parameter-efficient speech codec based on a cross-scale residual vector quantization scheme and transformers. Our model employs mirrored hierarchical window transformer blocks and performs step-wise decoding from coarse-to-fine feature representations. To enhance bitrate efficiency, we propose a novel combination of vector quantization techniques along with a pre-training paradigm. Extensive experiments demonstrate that ESC can achieve high-fidelity speech reconstruction with significantly lower model complexity, making it a promising alternative to existing convolutional audio codecs.
\end{abstract}

\section{Introduction}
Recent advancements in deep learning have demonstrated the superiority of neural speech codecs over traditional ones, which rely on complex expert design and psycho-acoustic knowledge \cite{opus, evs}. Early efforts integrating deep generative models, such as WaveNet \cite{wavenet} and SampleRNN \cite{samplernn}, into audio codecs have delivered promising results. These models, acting as powerful decoders, synthesize high-quality speech from intermediate representations produced by traditional codecs \cite{wavenet-codec, samplernn-codec}. However, their auto-regressive nature of the decoding process often introduces significant inference latency, limiting their practical application. 

Alternatively, some end-to-end neural audio codecs leverage the vector quantization (VQ) network first introduced by \citet{vqvae}. VQ networks use a learnable collection of code-vectors, known as a codebook, to quantize continuous vectors by assigning them to the nearest codeword. This discretization positions VQNs well-suited for both generation and compression tasks. Following this approach, existing VQ codecs \cite{soundstream, encodec, dac} typically employ a three-stage architecture: a convolutional encoder and decoder, and a residual vector quantization (RVQ) module \cite{rvq} applied in the latent space. The encoder and decoder downsample and upsample audio waveform features, creating hierarchical representations. RVQ further refines vanilla vector quantization by minimizing quantization error through a series of codebooks that recursively quantize the residuals from previous stages. Additionally, these codecs employ adversarial discriminators to remove artifacts and produce high-fidelity audio reconstructions. Substantial effort has been dedicated to designing effective audio discriminators, including an improved feature matching loss \cite{melgan}, as well as various multi-resolution waveform and spectrogram discriminators \cite{hifigan, soundstream, encodec, bigvgan_periodic_act}. VQ-based audio codecs have demonstrated remarkable performance in audio reconstruction, even at ultra-low bitrates.   

Despite these advantages, we find that convolutional VQ codecs heavily depend on powerful discriminators to produce high-quality audio, posing additional optimization challenges due to adversarial training. Moreover, these codecs tend to confront computational constraints, as they require a large number of parameters to balance high compression rates and reconstruction performance. To address these issues, our work develops a more parameter-efficient speech codec by reducing model complexity and implementing the following architectural improvements: 1) replacing convolutional layers with efficient Swin-Transformer Blocks (STBs) \cite{swin}, which can better model acoustic features; 2) utilizing the cross-scale residual vector quantization (CS-RVQ) scheme \cite{csvq} instead of RVQ, extending quantization from a fixed level to multiple levels.

In addition, training VQ codecs frequently leads to a significant challenge: \textit{codebook collapse}, where a fraction of the codebook remains underutilized in representing input vectors. This issue is frequently observed when training visual tokenizers for generative vision tasks \cite{sq-vae, regularized_vq, straightening}. To address this problem in speech compression, we propose combining product vector quantization (PVQ) \cite{gvq}, code factorization \cite{vq_projection}, and Euclidean normalization \cite{kmeans} to enhance codebook utilization. Furthermore, we introduce a learning paradigm to facilitate optimization, which includes a pre-training stage where the codebooks are deactivated and trained subsequently. 

In summary, the key contributions of our work are as follows:
\begin{itemize}
\item We introduce ESC, a fully transformer-based speech codec with cross-scale quantization structures. It achieves a superior tradeoff between compression rate, reconstruction quality, and model complexity, outperforming current state-of-the-art models. 
\item We propose a novel combination of vector quantization techniques within the cross-scale residual vector quantization (CS-RVQ) framework, coupled with a pre-training paradigm that effectively mitigates codebook collapse and enhances bitrate efficiency.  
\item Extensive comparisons with Descript's audio codec on a multilingual speech corpus demonstrate that transformers and CS-RVQ, the core components of ESC, are superior backbones for speech foundation models than the mainstream convolutions and RVQ.  
\end{itemize}

\section{Related Work}
\subsection{Neural Audio Codecs}
Recently, most notable neural audio codecs have been based on the vector quantization (VQ) network, including SoundStream \cite{soundstream}, EnCodec \cite{encodec}, and Descript's audio codec (DAC) \cite{dac}. SoundStream is distinguished as the first universal codec capable of handling diverse audio types. EnCodec improves compression rates by integrating a lightweight transformer language model within the discrete latent space and implements a streaming architecture. Building on similar backbones, \citet{dac} further explore the implications of quantization dropout, a technique for bitrate scalability, and demonstrate the superiority of periodic inductive bias functions over common activation functions for audio signal modeling \cite{bigvgan_periodic_act, snake_act}. These models directly process audio waveforms and are classified as time-domain codecs. 

In contrast, frequency-domain codecs focus on processing more intuitive audio spectrogram features. Lyra \cite{lyra}, for example, converts audio waveforms into log mel-spectrograms and directly quantizes them into tokens. Due to the non-invertibility of mel-spectrograms, it relies on a vocoder \cite{wavernn} for waveform synthesis. To circumvent the inefficiencies associated with heavy vocoders, some frequency-domain codecs, including TFNet \cite{tfnet} and our ESC, employ the invertible Short-time Fourier Transform (STFT) to convert waveforms into complex spectra. This design enables the reconstructed STFT spectra to be seamlessly inverted back into waveforms without information loss using inverse-STFT. Among recent audio codecs, DAC achieves state-of-the-art compression ratios and reconstruction quality, though its computation bottlenecks are sometimes overlooked. 

\subsection{Swin Transformers}
Vision Transformers (ViTs) \cite{vit} have outperformed convolutional neural networks (CNNs) in various image processing tasks, largely due to their superior ability to capture complex patterns. The Swin Transformer \cite{swin}, a notable variant, enhances this capability by employing a hierarchical approach with shifted window attention mechanisms, enabling it to scale efficiently to high-resolution signals while maintaining computational efficiency. In the context of image compression, Swin Transformers have demonstrated exceptional performance. Studies by \citet{transformer_compression} and \citet{devil_in_detail} show that Swin Transformers surpass CNNs in modeling spatial hierarchies and long-range dependencies. The attention mechanism facilitates the accurate preservation of essential details and textures, even at lower bitrates. These capabilities suggest that transformers could also be effective in applications beyond image compression, such as modeling audio spectrograms. 

\subsection{Vector Quantization}
In the Vector Quantized Variational Autoencoder (VQ-VAE) \citep{vqvae}, vector quantization (VQ) functions as a trainable layer that deterministically quantizes encoded latent variables by mapping them to their nearest neighbors in an embedding codebook. A VQ layer, denoted as $Q(\cdot \ ; \mathcal{C})$, is parameterized by a collection of continuous vectors $\mathcal{C} = \{\bm{c}_1,...,\bm{c}_K\}$, each referred to as a codeword, with its associated index known as a code. The layer quantizes a vector $\bm{z}_e \in \mathbb{R}^d$ to $\bm{z}_q \in \mathbb{R}^d$ by selecting the Euclidean nearest codeword $\bm{c}_k$ from the codebook $\mathcal{C}$, i.e.,
\begin{align}
\bm{z}_q := \bm{c}_k = \argmin_{\bm{c}_j \in \mathcal{C}} ||\bm{z}_e - \bm{c}_j||_2^2.
\end{align}
For convenience, we denote the output of the VQ function as $(\bm{z}_q, \tilde{z}_q)$, where $\tilde{z}_q := k$ represents the discrete code corresponding to the nearest codeword. During compression, the encoding process outputs the discrete index $\tilde{z}_q$, which is stored with a $\log_2 K$ bit budget. The decoding process starts by retrieving the continuous vector $\bm{z}_q$ from the codebook using the index $\tilde{z}_q$. The VQ function $Q(\cdot; \mathcal{C})$ is non-differentiable due to the $\argmin$ operator. Common strategies use a straight-through estimator (STE) \cite{ste} to bypass this in back-propagation. In other words, the gradient component $\frac{\partial \bm{z}_q}{\partial \bm{z}e}$ is estimated by identity. Additionally, auxiliary losses including codebook loss and commitment loss are proposed to pull the codewords and latent features closer:
\begin{align}
\mathcal{L}_{vq} = || \text{sg}(\bm{z}_e) - \bm{z}_q||_2^2 + \beta ||\bm{z}_e - \text{sg}(\bm{z}_q)||_2^2. \label{vq_loss}
\end{align} 
Here $\text{sg}(\cdot)$ denotes the stop-gradient operator. The first term updates the codebook with an $l_2$ error, pushing the codewords towards the input vectors. The second term ensures that $\bm{z}_{e}$ commits to the embedding without growing arbitrarily. The scalar $\beta$ balances the importance of updating the codebook and the encoder.

\subsection{Codebook Collapse}
Straight-through estimators (STEs) can lead to significant issues, most notably \textit{codebook collapse}, as detailed by \citet{wasserstein-vq}. In a recent study, \citet{straightening} provide a plausible explanation, attributing the collapse to an internal codebook covariate shift during training. Frequent adjustments in encoder representations cause misalignment with the codebook, resulting in only a subset of codewords being updated. Consequently, VQ layers are prone to divergence, often ending up with a significant number of inactive vectors. Various strategies have been proposed in generative modeling context to address this issue, including stochastic quantization \cite{sq-vae, regularized_vq}, self-annealed soft-to-hard quantization \cite{soft-to-hard-vq}, re-initializing codewords using K-means centroids every few epochs \cite{kmeans, jukebox}, and reformulating with finite scalar quantization \cite{fsq}. In audio compression, \citet{dac} address codebook collapse by down-projecting codewords \cite{vq_projection} and normalizing them within a Euclidean ball \cite{kmeans}.

\begin{figure*}[t]
    \centering
    \includegraphics[width=1.\textwidth]{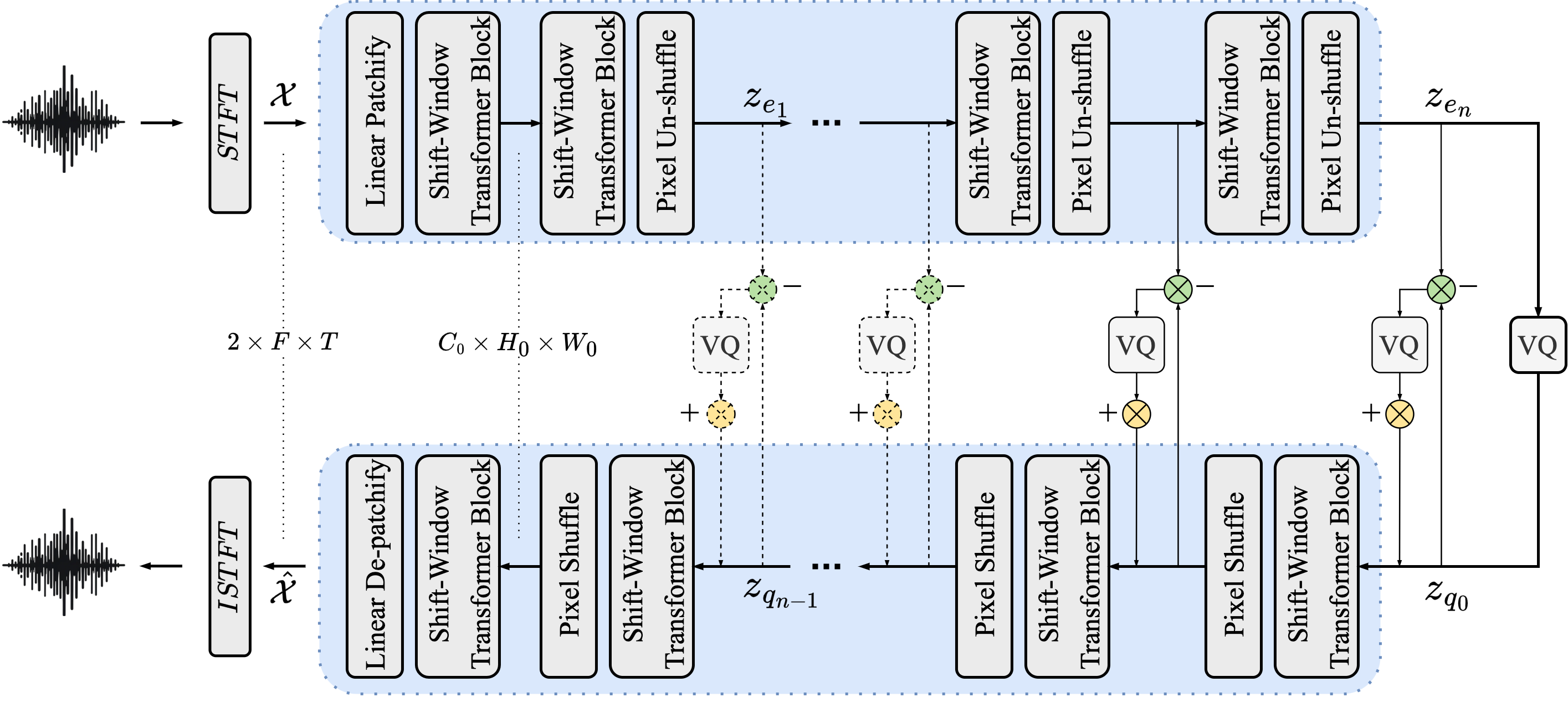}
    \caption{The framework of ESC: input speech is transformed to a complex STFT $\mathcal{X}$ and linearly embedded into patches. Encoder STBs iteratively halve the frequency resolution and produce hierarchical feature representations. Mirrored decoder STBs recover the frequency resolution by progressively leveraging coarse-to-fine quantized residual features between encoder and decoder hidden states. The entire network is solely composed of efficient transformer blocks and vector quantization layers. The figure displays a scenario when the deepest $3$ of $n+1$ total bitstreams (solid lines) are transmitted, with others left inactive.}
    \label{fig:architecture}
\end{figure*}

\section{Efficient Speech Codec (ESC)}
\subsection{Overall Architecture}
As illustrated in Figure~\ref{fig:architecture}, ESC operates on the complex spectrum $\mathcal{X} \in \mathbb{R}^{2 \times F \times T}$ derived from the Short-Time Fourier Transform (STFT) of a speech signal. Here, the real and imaginary components of $\mathcal{X}$ are treated as separate channels. Instead of using strided convolutions, ESC comprises a series of mirrored transformer encoder and decoder layers, each performing downsampling or upsampling to create coarse and fine representations, as described in Section~\ref{sec:swinT}. Starting from the quantized latents at the bottleneck VQ, the decoder progressively reconstructs the original spectrum by leveraging multi-level quantized residuals between the intermediate features of the encoder and decoder. This cross-scale decoding mechanism is further detailed in Section~\ref{sec:csrvq}. Finally, the reconstructed spectrum $\hat{\mathcal{X}}$ is transformed back into a waveform through the inverse-STFT.

\subsection{Notations}
We first define some notations for clarity. The encoder and decoder are denoted by $F_\phi(\cdot)$ and $G_\psi(\cdot)$, respectively, each being a composition of individual layer functions $f_{\phi_1}, \ldots, f_{\phi_n}$ and $g_{\psi_1}, \ldots, g_{\psi_n}$. We use $\mathcal{Z} \in \mathbb{R}^{C\times F\times T}$ to denote a spectrum feature and $\bm{z} \in \mathbb{R}^{CF} $ to denote a flattened time frame vector in $\mathcal{Z}$. Specifically, $\mathcal{Z}_{e_i}$ refers to the feature after the $i$-th encoder layer, and $\mathcal{Z}_{q_i}$ denotes the $i$-th decoder feature.
\begin{align}
    &\mathcal{Z}_{e_i} = f_{\phi_i} \circ ... \circ f_{\phi_1} \circ \mathcal{Z}_{e_0} \\
    &\mathcal{Z}_{q_i} = g_{\psi_i} \circ ... \circ g_{\psi_1} \circ \mathcal{Z}_{q_0},
\end{align} Here, $\mathcal{Z}_{e_0}$ is the original input feature and $\mathcal{Z}_{q_0}$ is the latent representation at the bottleneck.  

\subsection{Transformer Encoder and Decoder}\label{sec:swinT}
To effectively capture redundancies within audio signals, we replace convolutional layers with hierarchical Swin Transformer blocks (STBs) and their extended decoding counterparts. \\ 
\textbf{Patchify. } The encoder starts with a linear patchify module, where the complex spectrum $\mathcal{X}$ is divided into small patches and linearly up-projected: 
\begin{align}\mathcal{X} \in \mathbb{R}^{2 \times F \times T} \xrightarrow[]{\text{Patchify}} \mathcal{Z}_{e_0} \in \mathbb{R}^{C_0 \times H_0 \times W_0}.
\end{align} 
Here, the patch size across the frequency and temporal dimensions is $(\frac{F}{H_0}, \frac{T}{W_0})$. This step reduces the input resolution to alleviate the computational burden on attention computation. At the end of the decoder, a symmetric de-patchify module reshapes the decoded patch feature $\mathcal{Z}_{q_n}$ and linearly down-projects it to produce a recovered spectrum $\hat{\mathcal{X}}$.\\
\textbf{Swin Transformer blocks. } STBs in both the encoder and decoder employ window-based multi-head self-attention (W-MSA), partitioning spectrum features into smaller windows and computing attention in parallel within each window. This approach enables more efficient computation compared to vanilla attention mechanisms. To ensure connections between windows, STBs cascade two interleaved W-MSAs, with the outputs of the first being shifted for the second. This design allows STBs to capture local and global feature dependencies both effectively and efficiently. \\
\textbf{Downsampling and upsampling. } ESC maintains temporal resolution while scaling frequency resolution to equalize bitrates across different bitstreams. To achieve this, we modify the original patch merging/splitting modules with a single-dimensional pixel unshuffle/shuffle module \cite{pixelshuffle} along the frequency dimension. During encoder downsampling, an intermediate encoder spectrum feature $\mathcal{Z}_{e_i} \in \mathbb{R}^{C_i \times H_i \times W_i}$ is first reshaped and then projected by $P_{e_i} \in \mathbb{R}^{vC_{i} \times C_{i+1}}$ as follows:  
\begin{flalign}
    \xrightarrow[]{\text{reshape}}
    \mathbb{R}^{vC_i \times \frac{H_i}{v} \times W_i} 
    \xrightarrow[]{\text{proj}}
    \mathbb{R}^{C_{i+1} \times \frac{H_i}{v} \times W_i},
\end{flalign} where $v$ is the down-scaling factor. The upsampling process mirrors this operation in reverse. An intermediate decoder feature $\mathcal{Z}_{q_i} \in \mathbb{R}^{C_i \times H_i \times W_i}$ is first projected by $P_{q_i} \in \mathbb{R}^{C_i \times vC_{i+1}}$ and then reshaped, resulting in an up-scaled frequency resolution:
\begin{flalign}
    \xrightarrow[]{\text{proj}}
    \mathbb{R}^{vC_{i+1} \times H_i \times W_i}
    \xrightarrow[]{\text{reshape}}
    \mathbb{R}^{C_{i+1} \times vH_i \times W_i} .
\end{flalign}
Overall, the transformer encoder and decoder layers are mirrored, creating symmetric and hierarchical representations of the input audio spectrum. With these backbones, ESC is a fully transformer-based codec without any convolutional modules.  

\subsection{Cross-Scale Residual Vector Quantization}\label{sec:csrvq}
To achieve parameter-efficient modeling of audio signals, ESC employs multi-scale features that capture coarse-to-fine information. It integrates the more intuitive residual-based cross-scale vector quantization (CS-RVQ) framework proposed by \citet{csvq}, eliminating the need for additional networks to merge encoder and decoder features for improved reconstruction quality. As depicted in Algorithm~\ref{alg:cs-rvq-enc}, Algorithm~\ref{alg:cs-rvq-dec} and Figure~\ref{fig:architecture}, the decoding process is conditioned on the encoded quantized residuals between encoder and decoder features from low-to-high resolution scales. This approach differs from the commonly used residual vector quantization scheme, which operates solely at the lowest scale, relying on the highest-level information while overlooking low-level details.\\
\textbf{Encoding. } The encoding process begins with the encoder $F_{\phi}(\cdot)$, creating multi-scale encoder features $\bm{z}_{e_1},...,\bm{z}_{e_n}$. $\bm{z}_{e_n}$ is first quantized by the bottleneck quantizer $Q_0$ to form the lowest bitstream. This represents the simplest case when the number of transmitted bitstream $s$ is set to 1, and CS-RVQ reduces to a fixed-scale VQ at the bottleneck. For higher bitstreams, the residual between symmetric encoder and decoder at higher resolutions, $\bm{z}_{e_{n-i+1}} - \bm{z}_{q_{i-1}}$, is quantized by $Q_{i}$. The quantized residual $\bm{q}_i$ is then added back to $\bm{z}_{q_{i-1}}$ and decoded by the subsequent decoder layer $g_{\phi_i}(\cdot)$, producing the next decoder feature $\bm{z}_{q_{i}}$. Recursively, residuals at higher resolutions are progressively quantized, forming the remaining bitstreams (see Algorithm~\ref{alg:cs-rvq-enc}, Lines 3-6). This mechanism enables multi-scale learning, allowing the decoder layers to incrementally reduce quantization errors by conditioning on encoder-decoder residual features. When $s > 2$, this encoding process requires forward passing $s-2$ additional decoder layers to produce residuals at higher levels. After encoding, the input $\bm{z}_{e_0}$ is compressed into multi-level codes $\tilde{z}_{q_0}, \tilde{z}_{q_1}, \ldots, \tilde{z}_{q_{s-1}}$.
\begin{algorithm}[t]
\caption{CS-RVQ Encoding}\label{alg:cs-rvq-enc}
\begin{algorithmic}[1]
\Require A flattened time frame $\bm{z}_{e_0} \in \mathbb{R}^{C_0H_0}$, encoder $F_{\phi}(\cdot)$, decoder $G_{\psi}(\cdot)$, vector quantizers $Q_0,Q_1,...,Q_{s-1}$, number of bitstreams $s$

\State $\bm{z}_{e_1},...,\bm{z}_{e_n} \leftarrow F_{\phi}(\bm{z}_{e_0})$  \Comment{\small Encoder forward pass}
\normalsize
\State $\bm{z}_{q_0}, \tilde{z}_{q_0} \leftarrow Q_0(\bm{z}_{e_n})$ \Comment{\small bottom VQ}
\normalsize
\For{$i=1\dots s-2$}
  \State $\bm{q}_{i}, \tilde{z}_{{q}_{i}}  \leftarrow Q_{i}(\bm{z}_{e_{n-i+1}} - \bm{z}_{{q}_{i-1}})$ \label{enc-quant-residual}
  \State $\bm{z}_{{q}_{i}} \leftarrow g_{\psi_{i}}(\bm{z}_{{q}_{i-1}} + \bm{q}_{i})$
\EndFor \Comment{\small Encoding involves $s-2$ decoder layers}
\normalsize
\If{$s > 1$}
\State $\bm{q}_{s-1}, \tilde{z}_{{q}_{s-1}}  \leftarrow Q_{i}(\bm{z}_{e_{n-s+2}} \mathrel{-} \bm{z}_{{q}_{s-2}})$
\EndIf
\State \Return $\tilde{z}_{q_0}, \tilde{z}_{q_1}, ..., \tilde{z}_{q_{s-1}}$
\end{algorithmic}
\end{algorithm}

\begin{algorithm}[t]
\caption{CS-RVQ Decoding}\label{alg:cs-rvq-dec}
\begin{algorithmic}[1]
\Require Codes $\tilde{z}_{q_0}, \tilde{z}_{q_1}, ..., \tilde{z}_{q_{s-1}}$, decoder $G_{\psi}(\cdot)$, vector quantizers $Q_0,Q_1,...,Q_{s-1}$
\State $\bm{z}_{q_0} \xleftarrow[]{Q_0} \tilde{z}_{q_0}$ \Comment{\small Retrieve codewords from bottom VQ}
\normalsize
\For{$i=1\dots s-1$}
  \State $\bm{q}_{i} \xleftarrow[]{Q_{i}} \tilde{z}_{q_{i}}$ 
  \State $\bm{z}_{{q}_{i}} \leftarrow g_{\psi_{i}}(\bm{z}_{{q}_{i-1}} + \bm{q}_{i}) $ \label{dec-residual-refine}
\EndFor \Comment{\small Decoding refined by quantized residuals}
\normalsize
\For{$i=s \dots n$}
  \State $\bm{z}_{{q}_{i}} \leftarrow g_{\psi_{i}}(\bm{z}_{{q}_{i-1}})$
\EndFor \Comment{\small Continue with regular decoding}
\normalsize 
\State \Return $\bm{z}_{q_n}$

\end{algorithmic}
\end{algorithm} 
\noindent\textbf{Decoding. } The decoding process starts by retrieving the quantized latent at the bottom VQ using code $\tilde{z}_{q_0}$, which provides the initial decoder input $\bm{z}_{q_0}$. At higher levels, the codes $\tilde{z}_{q_1}, ..., \tilde{z}_{q_{s-1}}$ are iteratively used to retrieve codewords, producing multi-scale low-to-high quantized residuals $\bm{q}_1,...,\bm{q}_{s-1}$. In Algorithm~\ref{alg:cs-rvq-dec}, Lines 2-5, each quantized residual $\bm{q}_i$ is added back to the corresponding decoder feature $\bm{z}_{q_{i-1}}$ to refine the decoding process. Starting from the $s$-th decoder layer, there are no quantized residuals, and the remaining layers perform regular decoding. Finally, the recovered frame vector $\bm{z}_{q_n}$ is obtained, benefiting from $s-1$ quantized residual features. \\
\textbf{Training. } During training, the encoding and decoding processes are concatenated to form a complete forward pass. To enable bitrate scalability, we sample $s \sim \text{Uniform}\{1, \ldots, n\}$ at a rate $p$ within each training mini-batch. $p$ is a hyperparameter that balances the reconstruction quality at different bitrates, as proposed by \citet{dac}.

\subsection{Mitigating Codebook Collapse}
ESC performs a per-frame vector quantization. Before nearest neighbor searching, each input spectrum frame feature in $\mathcal{Z}$ needs to be flattened, merging the frequency and channel dimensions. This approach can result in large input vector dimensions for VQ, increasing the optimization challenges associated with codebook underutilization. \\
\textbf{Vector quantization setups. } To optimize the codebooks effectively, we modify the vanilla VQ by combining product vector quantization with code-vector factorization at each bitstream. Specifically, a flattened $d$-dimensional frame vector $\bm{z}_{e_{i}}$ is split into a set of $l$ sub-vectors. Each sub-vector $\bm{z}_{e_i}^{(m)}$ is down-projected by $W_{\text{in}} \in \mathbb{R}^{\frac{d}{l} \times u}$, where $u \ll d$, and then quantized using an individual codebook $\mathcal{C}_m$. The selected code-vector is then up-projected by $W_{\text{out}} \in \mathbb{R}^{u \times \frac{d}{l}}$ to form ${\bm{z}}_{q_i}^{(m)}$:
\begin{align}
  & \bm{z}_{e_{i}} \equiv  \{ \bm{z}_{e_i}^{(m)} \mid  \bm{z}_{e_i}^{(m)} \in \mathbb{R}^{\frac{C_i H_i}{l}}, m=1,...,l\}, \\
  & {\bm{z}}_{q_i}^{(m)} = W_{\text{out}}^\top \argmin_{\bm{c}_j \in \mathcal{C}_m} || W_{\text{in}}^\top \bm{z}_{e_i}^{(m)} - \bm{c}_j||_2.
\end{align} Additionally, both the projected vector $W_{\text{in}}^\top \bm{z}_{e_i}^{(m)}$ and codebook $\mathcal{C}_m$ are $l_2$ normalized before computing the distance matrix. This equalizes the scales of input vectors and codewords, enhancing codebook optimization by allowing a larger subset of codewords to receive gradients \cite{kmeans}.\\
\textbf{Pre-training paradigm. } Training transformers can be challenging, and jointly training them with VQ layers is even more difficult. To address this, we propose a pre-training paradigm that includes a warm-start to facilitate the learning process. Initially, all VQ layers are deactivated, meaning no quantization occurs. During this "pre-training" stage, only the encoder and decoder are updated within the CS-RVQ framework, allowing latent features to bypass the quantizers and flow directly into the decoder layers. Once the encoder and decoder have converged by minimizing reconstruction objectives, we resume training the entire VQ codec as usual. This approach helps mitigate the distribution shift of encoder representations by pre-optimizing the encoder. It helps stabilize codebook training and improve bitrate efficiency. Moreover, pre-training an auto-encoder is simpler, as it avoids the quantization errors associated with VQs. The detailed algorithm is provided in Appendix~\ref{appendix:pre-train}.

\subsection{Training Objectives}
To train our codec, we use a combination of reconstruction loss $\mathcal{L}_{recon}$ and vector quantization loss $\mathcal{L}_{vq}$. The reconstruction loss, $\mathcal{L}_{recon}$, consists of two components: an $l_2$ distance between the complex spectrum $\mathcal{X}$ and its reconstruction $\hat{\mathcal{X}}$, which forces the model to reconstruct the real and imaginary parts, weighted by $\lambda_1$, and a multi-scale mel-spectrogram loss \citep{dac}, weighted by $\lambda_2$. These are denoted as $\mathcal{L}_{stft}$ and $\mathcal{L}_{mel}$:
\begin{align}
  \mathcal{L}_{recon} = \lambda_1 \mathcal{L}_{mel} + \lambda_2 \mathcal{L}_{stft}.
\end{align} 
$\mathcal{L}_{vq}$ comprises the standard codebook and commitment losses as described in Equation~\ref{vq_loss}. It is averaged across the $l$ product vector quantizers and summed over all $s$ bitstreams. The final objective for joint optimization is the summation of $\mathcal{L}_{recon}$ and $\mathcal{L}_{vq}$. To deactivate the VQ layers during the pre-training stage, $\mathcal{L}_{vq}$ is set to zero.

\begin{figure*}[t]
  \centering
  \includegraphics[width=.95\linewidth]{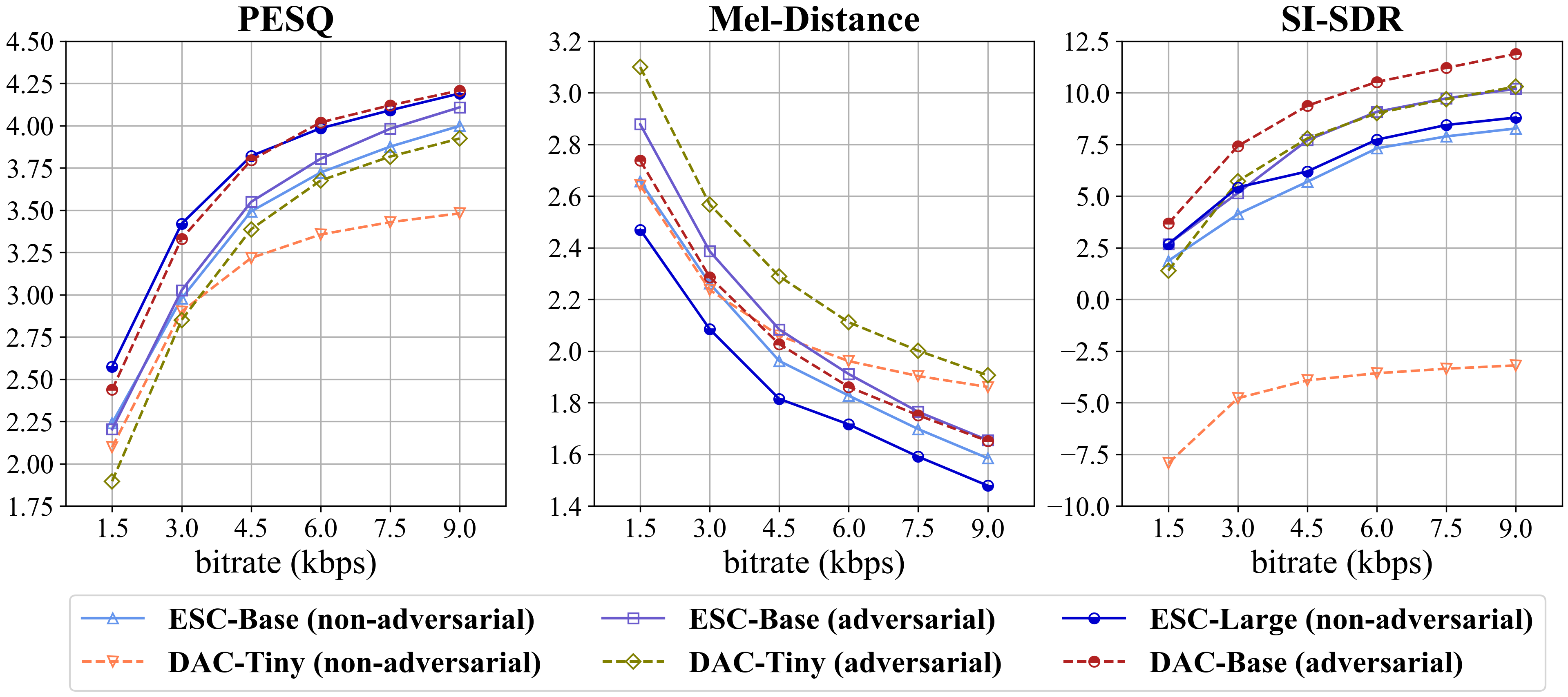}
   \caption{Reconstruction quality evaluation of different baseline codecs: dashed lines represent DAC baselines and solid lines represent our ESC models, with x-axis being transmission bits per second and y-axis being PESQ $(\uparrow)$, Mel-Distance $(\downarrow)$ and SI-SDR$(\uparrow)$. The metrics are averaged over our composed 1158 10-second speech clips.}
   \label{fig:perf}
\end{figure*}

\section{Experiments}
\subsection{Experimental Setup}
\textbf{Datasets. } We extract 150 hours of 16kHz multilingual clean speech from the DNS Challenge dataset \cite{dns2021}. Training samples are clipped into 3-second segments, and validation samples into 10-second segments. For evaluation, we compile 1158 multilingual 10-second speech clips with non-overlapping speakers from the LibriSpeech \cite{librispeech}, Multilingual LibriSpeech \cite{multi-librispeech}, and AIShell \cite{aishell-3} datasets. \\
\textbf{Baselines. } We compare our ESC against the current state-of-the-art time-domain codec DAC, by reproducing three versions\footnote{Reproduction settings are detailed in Appendix~\ref{sec:appendix-dac}.} on our dataset:\\
\text{1) DAC-Base (adversarial):} Descript's original released codec, operating on 16kHz audio signals. It has 74M parameter count in total. Its associated discriminator has 42M additional parameter count. \\ 
\text{2) DAC-Tiny (adversarial):} A smaller version of DAC-Base, with reduced encoder and decoder dimensions, for a fair comparison with ESC.\\
\text{3) DAC-Tiny (non-adversarial):} A smaller and non-adversarial version of DAC to assess the impact of discriminators on improving audio fidelity.\\
\textbf{Implementation details. } Similar to DAC baselines, we provide different versions of ESC\footnote{Complete configurations are detailed in Appendix~\ref{sec:appendix-esc}.}:\\
\text{1) ESC-Base (non-adversarial):} A base version codec consisting of 6 encoder/decoder layers, with bitrates ranging from 1.5 to 9.0 kbps. It contains 8.39M parameters when operating at 9.0 kbps.\\
\text{2) ESC-Base (adversarial):} An adversarial version using the same multi-scale multi-band waveform and spectrogram discriminator in DAC.\\
\text{3) ESC-Large (non-adversarial):} A scaled-up version with increased Swin Transformer layer depth, having 15.58M parameters at 9.0 kbps.\\
Our ESC variants are trained using the AdamW optimizer \cite{loshchilov2017decoupled} with a learning rate of 1e-4 and a weight decay of 1e-2. Training runs up to 0.4 million iterations without learning rate schedulers. The proposed pre-training phase consists of 0.75 million iterations. After pre-training, the codebooks are initialized with a Kaiming normalization distribution \cite{kaimingnormal}. The quantization dropout rate $p$ is set to 0.75. Loss weighting hyperparameters are set as $\lambda_1=0.25$, $\lambda_2=1.0$, and the commitment loss weighting $\beta=0.25$. For ESC-Base (adversarial), the $\mathcal{L}_{stft}$ component is eliminated. We use the HingeGAN \cite{lim2017geometric} adversarial loss formulation and the $l_1$ feature matching loss \cite{melgan}, following the approach of DAC. \\
\textbf{Automatic evaluation metrics. } We use objective metrics to efficiently evaluate reconstruction performance. These include the PESQ score \cite{pesq} from the speech enhancement domain, following \citet{csvq}; the $l_1$ distance between log mel-spectrograms of reference and decoded waveforms (Mel-Distance) \cite{dac}; and the scale-invariant source-to-distortion ratio (SI-SDR) \cite{sisdr}. To measure codec inference latency, we use the real-time factor (RTF), defined as the ratio of speech audio duration to model processing time \cite{encodec}.
\begin{table}[t]
\small
\resizebox{\columnwidth}{!}{
\begin{tabular}{ccccc}
\toprule
          &              &         & \multicolumn{2}{l}{Real Time Factor $\uparrow$} \\
          \cmidrule{4-5}
Codec     & Bitrate & \#Param. & Enc.          & Dec.         \\
\midrule
\multirow{3}{*}{ESC-Base}  & 3.0 kbps & 8.10M & 33.66 & 34.97 \\
                           & 6.0 kbps & 8.21M & 27.84 & 33.02 \\
                           & 9.0 kbps & 8.39M & 24.45 & 33.95 \\
\midrule
\multirow{3}{*}{DAC-Tiny}  & 3.0 kbps & 7.96M & 42.26 & 49.52 \\
                           & 6.0 kbps & 8.07M & 44.66 & 48.63 \\
                           & 9.0 kbps & 8.17M & 43.00 & 49.10 \\
\midrule
\multirow{3}{*}{ESC-Large} & 3.0 kbps & 15.30M & 17.91 & 20.81 \\
                           & 6.0 kbps & 15.41M & 15.48 & 19.87 \\
                           & 9.0 kbps & 15.58M & 13.73 & 20.56 \\
\midrule
\multirow{3}{*}{DAC-Base} & 3.0 kbps & 73.99M & 12.77 & 3.36 \\
                          & 6.0 kbps & 74.15M & 11.43 & 3.13 \\
                          & 9.0 kbps & 74.31M & 11.81 & 3.25 \\
\bottomrule
\end{tabular}
}
\caption{Complexity evaluation results of different baseline codecs: RTFs are measured from 100 10-second speech clips on an Intel Xeon Platinum 8352V CPU.}
\label{tab:complexity}
\end{table}
\subsection{Comparison with DAC}
We provide a thorough comparison focusing on compression rate, reconstruction quality, and inference efficiency, as shown in Table~\ref{tab:complexity} and Figure~\ref{fig:perf}.\\
\textbf{Performance evaluation. } First, it is important to note that ESC-Base and DAC-Tiny are similar in model size, each with approximately 8 million trainable parameters. Our results show that ESC-Base consistently outperforms DAC-Tiny across all bitrates, even without an adversarial discriminator. In contrast, DAC-Tiny's reconstruction quality significantly drops without a discriminator in training, particularly in SI-SDR statistics. This indicates a heavy reliance of DAC models on GANs for maintaining high reconstruction quality. Notably, ESC-Base is compatible with the same convolution-based GAN discriminator used in DAC, as evidenced by its improved performance across all metric curves in its adversarial variant. Additionally, ESC-Large demonstrates that increasing ESC's model size can further enhance performance, with its PESQ curve matching that of DAC-Base, the top-performing and largest model. While DAC-Base achieves higher SI-SDR values, ESC-Large records a smaller Mel-Distance. Thus, we conclude that the two codecs achieve comparable performance, even though ESC-Large is trained without an adversarial discriminator.\\
\textbf{Complexity evaluation. } Despite their exceptional performance, Descript's top-performing codecs face significant computational challenges. This is evident from Table~\ref{tab:complexity}, where the decoding real-time factor (RTF) for DAC-Base is approximately 3.0, making it rather impractical for real-time applications. In contrast, our transformer-based ESC achieves much higher decoding RTFs (approximately 34), indicating superior computational efficiency. Although ESC-Base is not as fast as DAC-Tiny due to the overhead of attention computation, it offers substantially better speech reconstruction capabilities, striking a favorable balance between compression performance and computation latency. Future work could incorporate transformer speedup techniques such as FlashAttention \cite{dao2022flashattention} to further enhance ESC’s latency further. Moreover, following the CS-RVQ scheme, ESC possesses faster encoding speeds at lower bitrates---a capability not evidently found in DAC models.

These results suggest that our transformer-based codec, equipped with CS-RVQ, is a more parameter-efficient foundation model compared to time-domain convolutional counterparts. ESC is shown to be a more lightweight and effective neural speech codec, as ESC-Large achieves comparable performance to DAC-Base without the need for a powerful discriminator. Specifically, it boasts approximately $\times$4.8 smaller model size, $\times$1.4 faster encoding speed, and $\times$6.4 faster decoding speed.

\begin{table}[t]
\centering
\Huge
\resizebox{\columnwidth}{!}{
\begin{tabular}{cccccc}
\toprule
Method & Bitrate & PESQ $\uparrow$ & Mel dist. $\downarrow$ & SI-SDR $\uparrow$ & VQ util. $\uparrow$ \\
\midrule
\multirow{3}{*}{CNN + RVQ}   & 3.0 kbps & 2.71 & 2.82 & 0.57 & 96.8\%  \\
                             & 6.0 kbps & 2.93 & 2.69 & 1.03 & 98.2\% \\
                             & 9.0 kbps & 2.96 & 2.68 & 1.05 & 98.7\% \\
\midrule
\multirow{3}{*}{CNN + CS-RVQ} & 3.0 kbps  & 2.70 & 2.81 & 2.19 & 96.6\% \\
                              & 6.0 kbps  & 3.47 & 2.41 & 3.79 & 97.7\% \\
                              & 9.0 kbps  & 3.75 & 2.25 & 4.16 & 97.3\% \\
\midrule
\multirow{3}{*}{SwinT + RVQ}  & 3.0 kbps  & 2.97 & 2.22 & 0.77 & \textbf{98.1}\% \\
                              & 6.0 kbps  & 3.14 & 2.08 & 1.35 & \textbf{99.0}\% \\
                              & 9.0 kbps  & 3.16 & 2.07 & 1.39 & \textbf{99.2}\% \\
\midrule
\multirow{3}{*}{\makecell{ESC-Base \\ (SwinT + CS-RVQ)}} & 3.0 kbps  & {3.07} & \textbf{2.21} & \textbf{3.55} & 97.8\% \\
                                      & 6.0 kbps  & \textbf{3.73} & \textbf{1.80} & \textbf{4.74} & 98.3\% \\
                                      & 9.0 kbps  & \textbf{3.92} & \textbf{1.62} & \textbf{5.33} & 97.9\% \\
\midrule
\multirow{3}{*}{\makecell{ESC-Base \\ w/o Pre-training}} & 3.0 kbps  & \textbf{3.09} & 2.25 & 1.75 & 97.7\% \\
                                      & 6.0 kbps  & 3.53 & 1.97 & 2.87 & 98.1\% \\
                                      & 9.0 kbps  & 3.58 & 1.89 & 2.88 & 86.5\% \\
\bottomrule
\end{tabular}
}
\caption{Performance evaluation of different ablation models: results are obtained from the 1157 10-second speech clips in our test dataset.}
\label{tab:ablation}
\end{table}
\subsection{Ablation Study}
To investigate the effectiveness of the proposed components in ESC, we conducted thorough ablation experiments\footnote{Implementation setups are detailed in Appendix~\ref{sec:appendix-ablation}.} by training frequency-domain codecs operating on complex STFT spectra with different architectures. For fair comparisons, all other ablation models listed in Table~\ref{tab:ablation} have similar model sizes to ESC-Base.\\
\textbf{Swin Transformers and CNNs. } To demonstrate that transformers are superior auto-encoder backbones in neural speech coding, we focus on two pairs of experiments: CNN/SwinT + RVQ and CNN/SwinT + CS-RVQ. In these experiments, the channel dimensions of the CNN blocks are set to match the hidden dimensions of the Swin Transformer Blocks (STBs). The comparison, as shown in Table~\ref{tab:ablation}, reveals that transformer-based codecs consistently outperform CNN-based codecs across all performance metrics and bitrates, regardless of the quantization scheme used. \\
\textbf{CS-RVQ and RVQ. } Table~\ref{tab:ablation} highlights that CS-RVQ is a superior quantization scheme compared to RVQ, regardless of whether the backbone is CNN or STB. RVQ-based codecs hit performance bottlenecks, as adding more VQs does not improve audio quality (\emph{e.g.}, from 6.0 kbps to 9.0 kbps). However, codecs using the CS-RVQ scheme do not face such bottlenecks at higher bitrates and consistently outperform their RVQ counterparts. CS-RVQ is therefore a superior vector quantization framework that leverages multi-scale features effectively. \\
\textbf{Effect of pre-training paradigm. } To evaluate the efficacy of the pre-training stage, we conducted an experiment of ESC-Base w/o pre-training. We monitored the VQ utilization rate, calculated as the sum of entropy (in bits) divided by the maximum number of bits from all transmitted bitstreams. This metric reflects bitrate efficiency and the fraction of seldom-used codewords. The results indicate that models with pre-training achieve a near 1.0 utilization rate. However, ESC-Base w/o pre-training displays a lower utilization rate at 9.0 kbps, and its reconstruction performance is also inferior to that of the fully pre-trained ESC-Base. These findings suggest that the pre-training paradigm indeed helps avoid bitrate wastage and improve audio reconstruction quality.

\section{Conclusions}
In this paper, we introduce ESC, the first fully transformer-based neural speech foundation model designed for multilingual speech coding. ESC surpasses existing state-of-the-art time-domain VQ-based codecs in terms of complexity and achieves comparable compression performance without the need for a powerful adversarial discriminator. Our extensive evaluations demonstrate that the cross-scale residual vector quantization scheme and the Swin Transformer backbones are better suited for neural speech coding than the convolutional blocks and residual vector quantization utilized in mainstream codecs. Overall, our study suggests a promising direction for speech foundation models. Future research could focus on expanding multi-scale vector quantization techniques and investigating additional transformer variants optimized for speech signal modeling. 

\section{Limitations}
First, recent neural audio codecs are increasingly utilized in downstream generation tasks, where the codec acts as a foundation model to create discrete acoustic representations \cite{audiolm, audiogen, vocos, valle, funcodec}. These compressed representations, treated as acoustic tokens, are suitable for auto-regressive language modeling in generative tasks. However, our work does not explore this important aspect. A promising future direction would be to evaluate ESC in downstream applications such as speech synthesis and speech recognition. We anticipate that the cross-scale code representations learned from transformer backbones could offer advantages over the fixed-scale features of mainstream convolutional codecs in these tasks. 

Second, different automatic metrics for audio evaluation can produce inconsistent results, which is evidenced in our results. To further strengthen our conclusions, it is necessary to conduct subjective evaluations involving human evaluators, such as MUSHRA listening tests \cite{mushra}. Despite this limitation, we provide a collection of demo speech samples publicly available in our codebase, which we hope will help demonstrate ESC's performance and compensate for the absence of subjective metrics.

Besides, the primary focus of this work is to demonstrate the superiority of transformer and cross-scale frameworks over other mainstream methods, rather than to develop a production-ready codec like DAC or EnCodec. Nonetheless, given the scalability of transformers \cite{scalinglaw}, increasing the ESC model size and training it on larger and more diverse audio datasets also represent a promising direction for enhancing its practical applicability. 

Finally, as discussed in Section~\ref{sec:csrvq}, the cross-scale residual vector quantization (CS-RVQ) scheme requires the partial use of decoder layers during the encoding process, introducing additional latency as the bitrate increases. Similar to residual vector quantization, CS-RVQ requires careful sampling of the transmitted bitstream during training to achieve scalable bitrates within a single model. This sampling strategy can lead to performance trade-offs across different bitrates and may cause instability during training. Therefore, future research in speech foundational models could explore leveraging alternative recurrent structures \cite{toderici2017full, johnston2018improved, diao2020drasic} to improve coding scalability and address these challenges.


\bibliography{refs}

\begin{thebibliography}{55}
\providecommand{\natexlab}[1]{#1}

\bibitem[{Agustsson et~al.(2017)Agustsson, Mentzer, Tschannen, Cavigelli, Timofte, Benini, and Gool}]{soft-to-hard-vq}
Eirikur Agustsson, Fabian Mentzer, Michael Tschannen, Lukas Cavigelli, Radu Timofte, Luca Benini, and Luc~V Gool. 2017.
\newblock Soft-to-hard vector quantization for end-to-end learning compressible representations.
\newblock \emph{Advances in neural information processing systems}, 30.

\bibitem[{Baevski et~al.(2019)Baevski, Schneider, and Auli}]{gvq}
Alexei Baevski, Steffen Schneider, and Michael Auli. 2019.
\newblock vq-wav2vec: Self-supervised learning of discrete speech representations.
\newblock \emph{arXiv preprint arXiv:1910.05453}.

\bibitem[{Bengio et~al.(2013)Bengio, L{\'e}onard, and Courville}]{ste}
Yoshua Bengio, Nicholas L{\'e}onard, and Aaron Courville. 2013.
\newblock Estimating or propagating gradients through stochastic neurons for conditional computation.
\newblock \emph{arXiv preprint arXiv:1308.3432}.

\bibitem[{Borsos et~al.(2023)Borsos, Marinier, Vincent, Kharitonov, Pietquin, Sharifi, Roblek, Teboul, Grangier, Tagliasacchi et~al.}]{audiolm}
Zal{\'a}n Borsos, Rapha{\"e}l Marinier, Damien Vincent, Eugene Kharitonov, Olivier Pietquin, Matt Sharifi, Dominik Roblek, Olivier Teboul, David Grangier, Marco Tagliasacchi, et~al. 2023.
\newblock Audiolm: a language modeling approach to audio generation.
\newblock \emph{IEEE/ACM Transactions on Audio, Speech, and Language Processing}.

\bibitem[{Dao et~al.(2022)Dao, Fu, Ermon, Rudra, and R{\'e}}]{dao2022flashattention}
Tri Dao, Dan Fu, Stefano Ermon, Atri Rudra, and Christopher R{\'e}. 2022.
\newblock Flashattention: Fast and memory-efficient exact attention with io-awareness.
\newblock \emph{Advances in Neural Information Processing Systems}, 35:16344--16359.

\bibitem[{D{\'e}fossez et~al.(2023)D{\'e}fossez, Copet, Synnaeve, and Adi}]{encodec}
Alexandre D{\'e}fossez, Jade Copet, Gabriel Synnaeve, and Yossi Adi. 2023.
\newblock \href {https://openreview.net/forum?id=ivCd8z8zR2} {High fidelity neural audio compression}.
\newblock \emph{Transactions on Machine Learning Research}.
\newblock Featured Certification, Reproducibility Certification.

\bibitem[{Dhariwal et~al.(2020)Dhariwal, Jun, Payne, Kim, Radford, and Sutskever}]{jukebox}
Prafulla Dhariwal, Heewoo Jun, Christine Payne, Jong~Wook Kim, Alec Radford, and Ilya Sutskever. 2020.
\newblock Jukebox: A generative model for music.
\newblock \emph{arXiv preprint arXiv:2005.00341}.

\bibitem[{Diao et~al.(2020)Diao, Ding, and Tarokh}]{diao2020drasic}
Enmao Diao, Jie Ding, and Vahid Tarokh. 2020.
\newblock Drasic: Distributed recurrent autoencoder for scalable image compression.
\newblock In \emph{2020 Data Compression Conference (DCC)}, pages 3--12. IEEE.

\bibitem[{Dietz et~al.(2015)Dietz, Multrus, Eksler, Malenovsky, Norvell, Pobloth, Miao, Wang, Laaksonen, Vasilache et~al.}]{evs}
Martin Dietz, Markus Multrus, Vaclav Eksler, Vladimir Malenovsky, Erik Norvell, Harald Pobloth, Lei Miao, Zhe Wang, Lasse Laaksonen, Adriana Vasilache, et~al. 2015.
\newblock Overview of the evs codec architecture.
\newblock In \emph{2015 IEEE International Conference on Acoustics, Speech and Signal Processing (ICASSP)}, pages 5698--5702. IEEE.

\bibitem[{Dosovitskiy et~al.(2020)Dosovitskiy, Beyer, Kolesnikov, Weissenborn, Zhai, Unterthiner, Dehghani, Minderer, Heigold, Gelly et~al.}]{vit}
Alexey Dosovitskiy, Lucas Beyer, Alexander Kolesnikov, Dirk Weissenborn, Xiaohua Zhai, Thomas Unterthiner, Mostafa Dehghani, Matthias Minderer, Georg Heigold, Sylvain Gelly, et~al. 2020.
\newblock An image is worth 16x16 words: Transformers for image recognition at scale.
\newblock \emph{arXiv preprint arXiv:2010.11929}.

\bibitem[{Du et~al.(2024)Du, Zhang, Hu, and Zheng}]{funcodec}
Zhihao Du, Shiliang Zhang, Kai Hu, and Siqi Zheng. 2024.
\newblock Funcodec: A fundamental, reproducible and integrable open-source toolkit for neural speech codec.
\newblock In \emph{ICASSP 2024-2024 IEEE International Conference on Acoustics, Speech and Signal Processing (ICASSP)}, pages 591--595. IEEE.

\bibitem[{gil Lee et~al.(2023)gil Lee, Ping, Ginsburg, Catanzaro, and Yoon}]{bigvgan_periodic_act}
Sang gil Lee, Wei Ping, Boris Ginsburg, Bryan Catanzaro, and Sungroh Yoon. 2023.
\newblock \href {https://openreview.net/forum?id=iTtGCMDEzS_} {Big{VGAN}: A universal neural vocoder with large-scale training}.
\newblock In \emph{The Eleventh International Conference on Learning Representations}.

\bibitem[{He et~al.(2015)He, Zhang, Ren, and Sun}]{kaimingnormal}
Kaiming He, Xiangyu Zhang, Shaoqing Ren, and Jian Sun. 2015.
\newblock Delving deep into rectifiers: Surpassing human-level performance on imagenet classification.
\newblock In \emph{Proceedings of the IEEE international conference on computer vision}, pages 1026--1034.

\bibitem[{Huh et~al.(2023)Huh, Cheung, Agrawal, and Isola}]{straightening}
Minyoung Huh, Brian Cheung, Pulkit Agrawal, and Phillip Isola. 2023.
\newblock Straightening out the straight-through estimator: Overcoming optimization challenges in vector quantized networks.
\newblock In \emph{International Conference on Machine Learning}, pages 14096--14113. PMLR.

\bibitem[{Jiang et~al.(2022{\natexlab{a}})Jiang, Peng, Xue, Zhang, and Lu}]{csvq}
Xue Jiang, Xiulian Peng, Huaying Xue, Yuan Zhang, and Yan Lu. 2022{\natexlab{a}}.
\newblock \href {https://doi.org/10.21437/Interspeech.2022-10084} {Cross-scale vector quantization for scalable neural speech coding}.
\newblock In \emph{Interspeech 2022}, pages 4222--4226.

\bibitem[{Jiang et~al.(2022{\natexlab{b}})Jiang, Peng, Zheng, Xue, Zhang, and Lu}]{tfnet}
Xue Jiang, Xiulian Peng, Chengyu Zheng, Huaying Xue, Yuan Zhang, and Yan Lu. 2022{\natexlab{b}}.
\newblock End-to-end neural speech coding for real-time communications.
\newblock In \emph{ICASSP 2022-2022 IEEE International Conference on Acoustics, Speech and Signal Processing (ICASSP)}, pages 866--870.

\bibitem[{Johnston et~al.(2018)Johnston, Vincent, Minnen, Covell, Singh, Chinen, Hwang, Shor, and Toderici}]{johnston2018improved}
Nick Johnston, Damien Vincent, David Minnen, Michele Covell, Saurabh Singh, Troy Chinen, Sung~Jin Hwang, Joel Shor, and George Toderici. 2018.
\newblock Improved lossy image compression with priming and spatially adaptive bit rates for recurrent networks.
\newblock In \emph{Proceedings of the IEEE conference on computer vision and pattern recognition}, pages 4385--4393.

\bibitem[{Kalchbrenner et~al.(2018)Kalchbrenner, Elsen, Simonyan, Noury, Casagrande, Lockhart, Stimberg, Oord, Dieleman, and Kavukcuoglu}]{wavernn}
Nal Kalchbrenner, Erich Elsen, Karen Simonyan, Seb Noury, Norman Casagrande, Edward Lockhart, Florian Stimberg, Aaron Oord, Sander Dieleman, and Koray Kavukcuoglu. 2018.
\newblock Efficient neural audio synthesis.
\newblock In \emph{International Conference on Machine Learning}, pages 2410--2419. PMLR.

\bibitem[{Kaplan et~al.(2020)Kaplan, McCandlish, Henighan, Brown, Chess, Child, Gray, Radford, Wu, and Amodei}]{scalinglaw}
Jared Kaplan, Sam McCandlish, Tom Henighan, Tom~B Brown, Benjamin Chess, Rewon Child, Scott Gray, Alec Radford, Jeffrey Wu, and Dario Amodei. 2020.
\newblock Scaling laws for neural language models.
\newblock \emph{arXiv preprint arXiv:2001.08361}.

\bibitem[{Kleijn et~al.(2018)Kleijn, Lim, Luebs, Skoglund, Stimberg, Wang, and Walters}]{wavenet-codec}
W~Bastiaan Kleijn, Felicia~SC Lim, Alejandro Luebs, Jan Skoglund, Florian Stimberg, Quan Wang, and Thomas~C Walters. 2018.
\newblock Wavenet based low rate speech coding.
\newblock In \emph{2018 IEEE international conference on acoustics, speech and signal processing (ICASSP)}, pages 676--680. IEEE.

\bibitem[{Kleijn et~al.(2021)Kleijn, Storus, Chinen, Denton, Lim, Luebs, Skoglund, and Yeh}]{lyra}
W~Bastiaan Kleijn, Andrew Storus, Michael Chinen, Tom Denton, Felicia~SC Lim, Alejandro Luebs, Jan Skoglund, and Hengchin Yeh. 2021.
\newblock Generative speech coding with predictive variance regularization.
\newblock In \emph{ICASSP 2021-2021 IEEE International Conference on Acoustics, Speech and Signal Processing (ICASSP)}, pages 6478--6482. IEEE.

\bibitem[{Klejsa et~al.(2019)Klejsa, Hedelin, Zhou, Fejgin, and Villemoes}]{samplernn-codec}
Janusz Klejsa, Per Hedelin, Cong Zhou, Roy Fejgin, and Lars Villemoes. 2019.
\newblock High-quality speech coding with sample rnn.
\newblock In \emph{ICASSP 2019-2019 IEEE International Conference on Acoustics, Speech and Signal Processing (ICASSP)}, pages 7155--7159. IEEE.

\bibitem[{Kong et~al.(2020)Kong, Kim, and Bae}]{hifigan}
Jungil Kong, Jaehyeon Kim, and Jaekyoung Bae. 2020.
\newblock Hifi-gan: Generative adversarial networks for efficient and high fidelity speech synthesis.
\newblock \emph{Advances in neural information processing systems}, 33:17022--17033.

\bibitem[{Kreuk et~al.(2022)Kreuk, Synnaeve, Polyak, Singer, D{\'e}fossez, Copet, Parikh, Taigman, and Adi}]{audiogen}
Felix Kreuk, Gabriel Synnaeve, Adam Polyak, Uriel Singer, Alexandre D{\'e}fossez, Jade Copet, Devi Parikh, Yaniv Taigman, and Yossi Adi. 2022.
\newblock Audiogen: Textually guided audio generation.
\newblock \emph{arXiv preprint arXiv:2209.15352}.

\bibitem[{Kumar et~al.(2019)Kumar, Kumar, De~Boissiere, Gestin, Teoh, Sotelo, De~Brebisson, Bengio, and Courville}]{melgan}
Kundan Kumar, Rithesh Kumar, Thibault De~Boissiere, Lucas Gestin, Wei~Zhen Teoh, Jose Sotelo, Alexandre De~Brebisson, Yoshua Bengio, and Aaron~C Courville. 2019.
\newblock Melgan: Generative adversarial networks for conditional waveform synthesis.
\newblock \emph{Advances in neural information processing systems}, 32.

\bibitem[{Kumar et~al.(2023)Kumar, Seetharaman, Luebs, Kumar, and Kumar}]{dac}
Rithesh Kumar, Prem Seetharaman, Alejandro Luebs, Ishaan Kumar, and Kundan Kumar. 2023.
\newblock \href {https://openreview.net/forum?id=qjnl1QUnFA} {High-fidelity audio compression with improved {RVQGAN}}.
\newblock In \emph{Thirty-seventh Conference on Neural Information Processing Systems}.

\bibitem[{{\L}a{\'n}cucki et~al.(2020){\L}a{\'n}cucki, Chorowski, Sanchez, Marxer, Chen, Dolfing, Khurana, Alum{\"a}e, and Laurent}]{kmeans}
Adrian {\L}a{\'n}cucki, Jan Chorowski, Guillaume Sanchez, Ricard Marxer, Nanxin Chen, Hans~JGA Dolfing, Sameer Khurana, Tanel Alum{\"a}e, and Antoine Laurent. 2020.
\newblock Robust training of vector quantized bottleneck models.
\newblock In \emph{2020 International Joint Conference on Neural Networks (IJCNN)}, pages 1--7. IEEE.

\bibitem[{Le~Roux et~al.(2019)Le~Roux, Wisdom, Erdogan, and Hershey}]{sisdr}
Jonathan Le~Roux, Scott Wisdom, Hakan Erdogan, and John~R Hershey. 2019.
\newblock Sdr--half-baked or well done?
\newblock In \emph{ICASSP 2019-2019 IEEE International Conference on Acoustics, Speech and Signal Processing (ICASSP)}, pages 626--630. IEEE.

\bibitem[{Lim and Ye(2017)}]{lim2017geometric}
Jae~Hyun Lim and Jong~Chul Ye. 2017.
\newblock Geometric gan.
\newblock \emph{arXiv preprint arXiv:1705.02894}.

\bibitem[{Liu et~al.(2021)Liu, Lin, Cao, Hu, Wei, Zhang, Lin, and Guo}]{swin}
Ze~Liu, Yutong Lin, Yue Cao, Han Hu, Yixuan Wei, Zheng Zhang, Stephen Lin, and Baining Guo. 2021.
\newblock Swin transformer: Hierarchical vision transformer using shifted windows.
\newblock In \emph{Proceedings of the IEEE/CVF International Conference on Computer Vision (ICCV)}.

\bibitem[{Loshchilov(2017)}]{loshchilov2017decoupled}
I~Loshchilov. 2017.
\newblock Decoupled weight decay regularization.
\newblock \emph{arXiv preprint arXiv:1711.05101}.

\bibitem[{Mehri et~al.(2017)Mehri, Kumar, Gulrajani, Kumar, Jain, Sotelo, Courville, and Bengio}]{samplernn}
Soroush Mehri, Kundan Kumar, Ishaan Gulrajani, Rithesh Kumar, Shubham Jain, Jose Sotelo, Aaron Courville, and Yoshua Bengio. 2017.
\newblock \href {https://openreview.net/forum?id=SkxKPDv5xl} {Sample{RNN}: An unconditional end-to-end neural audio generation model}.
\newblock In \emph{International Conference on Learning Representations}.

\bibitem[{Mentzer et~al.(2024)Mentzer, Minnen, Agustsson, and Tschannen}]{fsq}
Fabian Mentzer, David Minnen, Eirikur Agustsson, and Michael Tschannen. 2024.
\newblock \href {https://openreview.net/forum?id=8ishA3LxN8} {Finite scalar quantization: {VQ}-{VAE} made simple}.
\newblock In \emph{The Twelfth International Conference on Learning Representations}.

\bibitem[{Oord et~al.(2016)Oord, Dieleman, Zen, Simonyan, Vinyals, Graves, Kalchbrenner, Senior, and Kavukcuoglu}]{wavenet}
Aaron van~den Oord, Sander Dieleman, Heiga Zen, Karen Simonyan, Oriol Vinyals, Alex Graves, Nal Kalchbrenner, Andrew Senior, and Koray Kavukcuoglu. 2016.
\newblock Wavenet: A generative model for raw audio.
\newblock \emph{arXiv preprint arXiv:1609.03499}.

\bibitem[{Panayotov et~al.(2015)Panayotov, Chen, Povey, and Khudanpur}]{librispeech}
Vassil Panayotov, Guoguo Chen, Daniel Povey, and Sanjeev Khudanpur. 2015.
\newblock Librispeech: an asr corpus based on public domain audio books.
\newblock In \emph{2015 IEEE international conference on acoustics, speech and signal processing (ICASSP)}, pages 5206--5210. IEEE.

\bibitem[{Pratap et~al.(2020)Pratap, Xu, Sriram, Synnaeve, and Collobert}]{multi-librispeech}
Vineel Pratap, Qiantong Xu, Anuroop Sriram, Gabriel Synnaeve, and Ronan Collobert. 2020.
\newblock Mls: A large-scale multilingual dataset for speech research.
\newblock \emph{arXiv preprint arXiv:2012.03411}.

\bibitem[{Reddy et~al.(2021)Reddy, Dubey, Gopal, Cutler, Braun, Gamper, Aichner, and Srinivasan}]{dns2021}
Chandan~KA Reddy, Harishchandra Dubey, Vishak Gopal, Ross Cutler, Sebastian Braun, Hannes Gamper, Robert Aichner, and Sriram Srinivasan. 2021.
\newblock Icassp 2021 deep noise suppression challenge.
\newblock In \emph{ICASSP 2021-2021 IEEE International Conference on Acoustics, Speech and Signal Processing (ICASSP)}, pages 6623--6627. IEEE.

\bibitem[{Series(2014)}]{mushra}
B~Series. 2014.
\newblock Method for the subjective assessment of intermediate quality level of audio systems.
\newblock \emph{International Telecommunication Union Radiocommunication Assembly}.

\bibitem[{Shi et~al.(2016)Shi, Caballero, Husz{\'a}r, Totz, Aitken, Bishop, Rueckert, and Wang}]{pixelshuffle}
Wenzhe Shi, Jose Caballero, Ferenc Husz{\'a}r, Johannes Totz, Andrew~P Aitken, Rob Bishop, Daniel Rueckert, and Zehan Wang. 2016.
\newblock Real-time single image and video super-resolution using an efficient sub-pixel convolutional neural network.
\newblock In \emph{Proceedings of the IEEE conference on computer vision and pattern recognition}, pages 1874--1883.

\bibitem[{Shi et~al.(2020)Shi, Bu, Xu, Zhang, and Li}]{aishell-3}
Yao Shi, Hui Bu, Xin Xu, Shaoji Zhang, and Ming Li. 2020.
\newblock Aishell-3: A multi-speaker mandarin tts corpus and the baselines.
\newblock \emph{arXiv preprint arXiv:2010.11567}.

\bibitem[{Siuzdak(2023)}]{vocos}
Hubert Siuzdak. 2023.
\newblock Vocos: Closing the gap between time-domain and fourier-based neural vocoders for high-quality audio synthesis.
\newblock \emph{arXiv preprint arXiv:2306.00814}.

\bibitem[{Takida et~al.(2022)Takida, Shibuya, Liao, Lai, Ohmura, Uesaka, Murata, Takahashi, Kumakura, and Mitsufuji}]{sq-vae}
Yuhta Takida, Takashi Shibuya, WeiHsiang Liao, Chieh-Hsin Lai, Junki Ohmura, Toshimitsu Uesaka, Naoki Murata, Shusuke Takahashi, Toshiyuki Kumakura, and Yuki Mitsufuji. 2022.
\newblock {SQ-VAE}: Variational bayes on discrete representation with self-annealed stochastic quantization.
\newblock In \emph{International Conference on Machine Learning}.

\bibitem[{Toderici et~al.(2017)Toderici, Vincent, Johnston, Jin~Hwang, Minnen, Shor, and Covell}]{toderici2017full}
George Toderici, Damien Vincent, Nick Johnston, Sung Jin~Hwang, David Minnen, Joel Shor, and Michele Covell. 2017.
\newblock Full resolution image compression with recurrent neural networks.
\newblock In \emph{Proceedings of the IEEE conference on Computer Vision and Pattern Recognition}, pages 5306--5314.

\bibitem[{Union(2007)}]{pesq}
IT~Union. 2007.
\newblock Wideband extension to recommendation p. 862 for the assessment of wideband telephone networks and speech codecs.
\newblock \emph{International Telecommunication Union, Recommendation P}, 862.

\bibitem[{Valin et~al.(2012)Valin, Vos, and Terriberry}]{opus}
Jean-Marc Valin, Koen Vos, and Timothy Terriberry. 2012.
\newblock Definition of the opus audio codec.
\newblock Technical report.

\bibitem[{Van Den~Oord et~al.(2017)Van Den~Oord, Vinyals et~al.}]{vqvae}
Aaron Van Den~Oord, Oriol Vinyals, et~al. 2017.
\newblock Neural discrete representation learning.
\newblock \emph{Advances in neural information processing systems}, 30.

\bibitem[{Vasuki and Vanathi(2006)}]{rvq}
A~Vasuki and PT~Vanathi. 2006.
\newblock A review of vector quantization techniques.
\newblock \emph{IEEE Potentials}, 25(4):39--47.

\bibitem[{Vuong et~al.(2023)Vuong, Le, Zhao, Zheng, Harandi, Cai, and Phung}]{wasserstein-vq}
Tung-Long Vuong, Trung Le, He~Zhao, Chuanxia Zheng, Mehrtash Harandi, Jianfei Cai, and Dinh Phung. 2023.
\newblock Vector quantized wasserstein auto-encoder.
\newblock \emph{arXiv preprint arXiv:2302.05917}.

\bibitem[{Wang et~al.(2023)Wang, Chen, Wu, Zhang, Zhou, Liu, Chen, Liu, Wang, Li et~al.}]{valle}
Chengyi Wang, Sanyuan Chen, Yu~Wu, Ziqiang Zhang, Long Zhou, Shujie Liu, Zhuo Chen, Yanqing Liu, Huaming Wang, Jinyu Li, et~al. 2023.
\newblock Neural codec language models are zero-shot text to speech synthesizers.
\newblock \emph{arXiv preprint arXiv:2301.02111}.

\bibitem[{Yu et~al.(2022)Yu, Li, Koh, Zhang, Pang, Qin, Ku, Xu, Baldridge, and Wu}]{vq_projection}
Jiahui Yu, Xin Li, Jing~Yu Koh, Han Zhang, Ruoming Pang, James Qin, Alexander Ku, Yuanzhong Xu, Jason Baldridge, and Yonghui Wu. 2022.
\newblock \href {https://openreview.net/forum?id=pfNyExj7z2} {Vector-quantized image modeling with improved {VQGAN}}.
\newblock In \emph{International Conference on Learning Representations}.

\bibitem[{Zeghidour et~al.(2021)Zeghidour, Luebs, Omran, Skoglund, and Tagliasacchi}]{soundstream}
Neil Zeghidour, Alejandro Luebs, Ahmed Omran, Jan Skoglund, and Marco Tagliasacchi. 2021.
\newblock Soundstream: An end-to-end neural audio codec.
\newblock \emph{IEEE/ACM Transactions on Audio, Speech, and Language Processing}, 30:495--507.

\bibitem[{Zhang et~al.(2023)Zhang, Zhan, Theobalt, and Lu}]{regularized_vq}
Jiahui Zhang, Fangneng Zhan, Christian Theobalt, and Shijian Lu. 2023.
\newblock Regularized vector quantization for tokenized image synthesis.
\newblock In \emph{Proceedings of the IEEE/CVF Conference on Computer Vision and Pattern Recognition}, pages 18467--18476.

\bibitem[{Zhu et~al.(2021)Zhu, Yang, and Cohen}]{transformer_compression}
Yinhao Zhu, Yang Yang, and Taco Cohen. 2021.
\newblock Transformer-based transform coding.
\newblock In \emph{International Conference on Learning Representations}.

\bibitem[{Ziyin et~al.(2020)Ziyin, Hartwig, and Ueda}]{snake_act}
Liu Ziyin, Tilman Hartwig, and Masahito Ueda. 2020.
\newblock Neural networks fail to learn periodic functions and how to fix it.
\newblock \emph{Advances in Neural Information Processing Systems}, 33:1583--1594.

\bibitem[{Zou et~al.(2022)Zou, Song, and Zhang}]{devil_in_detail}
Renjie Zou, Chunfeng Song, and Zhaoxiang Zhang. 2022.
\newblock The devil is in the details: Window-based attention for image compression.
\newblock In \emph{Proceedings of the IEEE/CVF conference on computer vision and pattern recognition}, pages 17492--17501.

\end{thebibliography}

\newpage
\appendix
\section{Pre-training Paradigm}\label{appendix:pre-train}
The proposed pre-training paradigm for optimizing vector quantization layers is detailed in Algorithm~\ref{alg:training}. During the pre-training phase, all vector quantization layers are bypassed, effectively reducing the codec to a standard autoencoder trained solely on reconstruction losses (Lines 2-4). Once the encoder and decoder reach a certain level of convergence, the VQ layers are reactivated, and joint optimization resumes. In the pre-training phase, we set the $\argmin$ nearest-neighbor selection as an identity function, making $\bm{z}_q$ equal to the input vector $\bm{z}_e$.

\begin{algorithm}[htbp]
\caption{Pre-training Paradigm}\label{alg:training}
\begin{algorithmic}[1]
\Repeat
  \State $\hat{\mathcal{X}} = 
  G_{\psi}(F_{\phi}(\mathcal{X}))$
  \State $\mathcal{L} = \mathcal{L}_{recon}(\mathcal{X}, \hat{\mathcal{X}})$ 
  \State take gradient descent step on $\nabla_{\phi}\mathcal{L},\nabla_{\psi}\mathcal{L}$
\Until converged
\State {activate VQs and continue learning as usual}
\end{algorithmic}
\end{algorithm}

\section{Experiment Details}
\subsection{DAC Reproduction Setups}
\label{sec:appendix-dac}
Our customized reproduction of DAC models closely follows the official development scripts. The original DAC model, designed for 16kHz audio signals, employs 12 VQ layers in its residual VQ module, supporting bitrates ranging from 0.5 kbps to 6.0 kbps. To ensure a fair comparison with ESC at similar bitrate levels, we extended the number of VQ layers in the RVQ module to 18, resulting in DAC-Base. For DAC-Tiny, we reduced the encoder dimension from 64 to 32 and the decoder dimension from 1536 to 288, while keeping other parameters unchanged. All DAC baselines were trained for 0.4 million iterations with a batch size of 16 on our multilingual speech dataset. Additional configuration details can be found in the official release\footnote{The official configuration for 16kHz DAC model is available at \url{https://github.com/descriptinc/descript-audio-codec/blob/main/conf/final/16khz.yml}}.

\subsection{ESC Architecture Configurations}
\label{sec:appendix-esc}
Overall, all three ESC variants are trained in distributed setups for 0.4 million iterations across 4 NVIDIA RTX 4090 GPUs with a total batch size of 36. These experiments took approximately 100 GPU hours. 

\subsubsection{Model Parameters}
The parameter configurations for ESC-Base are provided in Table~\ref{esc-base-param}. For STFT transformation, we use a 20 ms window length and a 5 ms hop length, implemented with torchaudio. The number of FFT points is set to 382, resulting in a frequency dimension of 192.  In the Swin Transformer, the layer depth represents the number of Swin Transformer Blocks (STBs) cascaded at each encoder and decoder layer. We use GELU activation functions and LayerNorm for normalization. In the down-sampling/up-sampling module, we use a scaling factor of $v=2$ to un-shuffle/shuffle along the frequency resolution only. Before the vector quantization layers, ESC processes two overlapping time frames together. To implement this, the flattened spectrum feature $\mathcal{Z}$ is reshaped from $\mathbb{R}^{W_i \times H_i C_i}$ to $\mathbb{R}^{W_i / 2 \times 2 H_i C_i}$. Each frame is then split into sub-vectors, down-projected, and $l_2$ normalized before computing the distance matrix. The VQ layer at each bitstream of ESC-Base consumes $\log_2 1024 \times 3 \times 150 = 4500$ bits per 3-second input speech (\emph{i.e.}, 1.5 kbps bitrate). For the scaled-up ESC-Large variant, we increase the STB layer depth from 2 to 4 while keeping the other configurations unchanged. 

\begin{table}[htbp]
\LARGE
\resizebox{\columnwidth}{!}{
\begin{tabular}{ccc}
\toprule
Modules & Parameters                  & Values                          \\
\midrule
\multirow{2}{*}{STFT} & Window/Hop Length                & [20ms, 5ms]                     \\
& Number of FFT                    & 382                             \\
\midrule
\multirow{5}{*}{Encoder/Decoder} & Patch Size                       & [3, 2]                          \\
& Layer Dims $C_1, ..., C_6$ & [45, 72, 96, 144, 192, 384] \\
& Attention Heads                  & [3, 3, 6, 12, 24, 24]       \\
& Layer Depth                        & 2                               \\
& Scaling Factor $v$                 & 2                               \\
\midrule
\multirow{3}{*}{Vector Quantization} & Product VQ Size $l$                     & 3                               \\
& Codevector Dimension $u$             & 8                               \\
& Codebook Size $K$            & 1024                            \\
\bottomrule
\end{tabular}
}
\caption{Parameter configurations of model variant ESC-Base, which comprises 6 encoder/decoder layers.}
\label{esc-base-param}
\end{table}

\subsubsection{Adversarial Training Setup}
In the ESC-Base (adversarial) variant, the GAN discriminator is identical to the one used in DAC, consisting of a multi-period discriminator (MPD), multi-band discriminator (MBD), and multi-scale STFT discriminator (MSD), totaling over 42 million parameters. The adversarial loss formulation follows the official DAC-Base (adversarial) configuration. Additionally, we maintain the pre-training paradigm for 0.75 million iterations in this variant, with the discriminator intervening in training only after the pre-training stage finishes.

\subsection{Details on Ablation Experiments}\label{sec:appendix-ablation}
All ablation models operate on the complex STFT spectrum, as in ESC (SwinT + CS-RVQ), using the same STFT configurations specified in Table~\ref{esc-base-param}. These models were trained for 0.25 million iterations, with 0.025 million iterations allocated for pre-training. The Swin Transformer configurations mirror those used in ESC-Base. Similarly, the vector quantization setup in the CS-RVQ models follows that of ESC-Base. In total, the ablation experiments required approximately 80 hours on 4 RTX 4090 GPUs.
\subsubsection{Convolution Blocks}
For models with CNN backbones, the convolutional channel dimensions were set to match the hidden sizes of the STB-based models. Each CNN block consists of one residual unit and one downsampling/upsampling 2D convolutional layer with a stride of 2 along the frequency resolution only. The residual unit consists of two 2D convolutional layers, each followed by BatchNorm and Parametric ReLU activation. 

\subsubsection{Residual Vector Quantization Setups}
For models using RVQs, we adapted the basic RVQ framework commonly used in time-domain codecs. To process frequency-domain spectrum features at the latent bottleneck, we combined RVQ with product vector quantization. Specifically, the flattened time frame vector is split into sub-group vectors, which are then recursively quantized, as in standard RVQs. We set the number of product VQs to 3 and the number of residual VQs to 6, ensuring the bitrate levels match those of ESC-Base (1.5 kbps per bitstream, 6 in total). 

\end{document}